\documentclass[reprint,showpacs,amsmath,amssymb,twocolumn,prl,superscriptaddress]{revtex4-1}

\usepackage[dvips]{graphicx} 
\usepackage{enumerate}
\usepackage{epsfig}
\usepackage{subfigure}
\usepackage{xcolor}
\usepackage[T1]{fontenc}
\usepackage[expert]{mathdesign}
\usepackage{fullpage}
\usepackage{amsthm,amsfonts,amssymb,amscd,mathrsfs,eufrak,xspace,framed}
\usepackage{amsmath}
\usepackage{color}
\usepackage{setspace}
\usepackage{url}
\usepackage{enumitem}
\usepackage{multirow}

\newcommand{\ket}[1]{\mbox{$\left| #1 \right\rangle$}}
\newcommand{\braket}[2]{\mbox{$\left\langle #1 | #2 \right\rangle$}}

\begin{document}

\title{Proof-of-principle experimental realization of a qubit-like qudit-based quantum key distribution scheme}

\author{Shuang Wang}
\author{Zhen-Qiang Yin}
\affiliation{CAS Key Laboratory of Quantum Information, University of Science and Technology of China, Hefei 230026, P. R. China\\
and Synergetic Innovation Center of Quantum Information $\&$ Quantum Physics, University of Science and Technology of China, Hefei, Anhui 230026, P. R. China,\\
State Key Laboratory of Cryptology, P. O. Box 5159, Beijing 100878, P. R. China}
\author{H.F. Chau}
\affiliation{Department of Physics and Center of Theoretical and Computational Physics, Pokfulam Road, Hong Kong}
\author{Wei Chen}
\email{weich@ustc.edu.cn}
\author{Chao Wang}
\author{Guang-Can Guo}
\author{Zheng-Fu Han}
\email{zfhan@ustc.edu.cn}
\affiliation{CAS Key Laboratory of Quantum Information, University of Science and Technology of China, Hefei 230026, P. R. China\\
and Synergetic Innovation Center of Quantum Information $\&$ Quantum Physics, University of Science and Technology of China, Hefei, Anhui 230026, P. R. China,\\
State Key Laboratory of Cryptology, P. O. Box 5159, Beijing 100878, P. R. China}

\begin{abstract}
In comparison to qubit-based protocols, qudit-based quantum key distribution (QKD) ones generally allow two cooperative parties to share unconditionally secure keys under a higher channel noise. However, it is very hard to prepare and measure the required quantum states in qudit-based protocols in general. One exception is the recently proposed highly error tolerant qudit-based protocol known as the Chau15 \cite{chau15}. Remarkably, the state preparation and measurement in this protocol can be done relatively easily since the required states are phase encoded almost like the diagonal basis states of a qubit. Here we report the first proof-of-principle demonstration of the Chau15 protocol. One highlight of our experiment is that its post-processing is based on practical one-way manner, while the original proposal in Ref.~\cite{chau15} relies on complicated two-way post-processing, which is a great challenge in experiment. In addition, by manipulating time-bin qudit and measurement with a variable delay interferometer, our realization is extensible to qudit with high-dimensionality and confirms the experimental feasibility of the Chau15 protocol.
\end{abstract}

\maketitle

{\it Introduction.} Quantum key distribution (QKD) allows two distant peers Alice and Bob to share secret key bits through a quantum channel which is accessed
by a malicious eavesdropper Eve \cite{Bennett:BB84:1984,Ekert:QKD:1991}. In a typical QKD protocol, Alice encodes random classical bits into quantum states, and sends them to Bob, who measures the incoming quantum states to decode Alice's classical bits. Then by classical communications and random sampling, Alice and Bob can obtain raw key bits, whose error rate can be also estimated. Finally, by classical post-processing, Alice and Bob can generate secret key bits. The most commonly used QKD protocol is the Bennett-Brassard 1984 (BB84) protocol \cite{Bennett:BB84:1984} combined with the decoy states method \cite{Hwang:Decoy:2003,Wang:Decoy:2005,Lo:Decoy:2005}. In the past decade, tremendous progresses in experimental decoy states BB84 have been achieved \cite{Zhao:DecoyExp:2006,Rosenberg:ExpDecoy:2007,Zeilinger:ExpDecoy:2007,Peng:ExpDecoy:2007, Zhao:Decoy60km:2006,Yuan:ExpDecoy2007,Frolich:QKDnet:2013}. Lately, several novel QKD protocols have been proposed \cite{Lo:MDIQKD:2012,Braunstein:MIQKD:2012,sasaki2014practical,chau15}. Among these protocols, measurement-device-independent (MDI)-QKD \cite{Lo:MDIQKD:2012} is immune to all detector-side-channel attacks and have been proven to be a feasible QKD scheme \cite{Rubenok:MIQKDexp:2013,liu:MIQKDexp:2013,daSilva:MIQKD:2013,Tang:MDIQKDexp:2013, RFIMDIQKDexp}. And the round-Robin-differential-phase-shift (RRDPS) protocol features that  monitoring signal disturbance can be bypassed \cite{sasaki2014practical}, thus becomes another hot topic. In the RRDPS protocol, Alice has to prepare trains of pulses each consisting of $L$ pulses.  In the simple case that each train contains only one photon, the secret key rate $R$ of the RRDPS protocol in the one-way classical communication setting equals $1-h_2(e_\text{bit})-h_2(1/(L-1))$, where $e_\text{bit}$ is the bit error rate of the raw key \cite{sasaki2014practical}.  In other words, the RRDPS protocol can tolerate much higher bit error rate of the raw key than the BB84 protocol when $L$ is sufficiently large. However, implementing the RRDPS protocol for large $L$ posts a great experimental challenge. In spite of the experimental difficulty, several experimental demonstrations of RRDPS protocol have been reported \cite{PassiveRRDPS,RRDPSexp1,RRDPSexp2,RRDPSexp3}.

 Inspired by the RRDPS protocol, Chau recently proposed a novel QKD protocol \cite{chau15}, called the Chau15 protocol, which can tolerate very high error rate and has a simpler implementation than the RRDPS protocol. In the Chau15 protocol, for each trial Alice randomly picks two distinct numbers $i,j$ from the set $\{1,\ldots,L\}$ and a raw key bit $k \in \{ 0 \equiv +, 1 \equiv - \}$, then prepares a quantum state $\ket{\psi^\pm_{ij}}=(\ket{i}\pm\ket{j})/\sqrt{2}$ according to the value of $k$ used.  That is to say, Alice encodes each raw bit in the phase between the two time bins $\ket{i}$ and $\ket{j}$. Alice sends $\ket{\psi^k_{ij}}$ to Bob, who then randomly picks two distinct numbers $m,n$ in $\{1,\ldots,L\}$ and measures the incoming photon along $\{\ket{\psi^\pm_{mn}}\}$. Bob records his raw key bit as $0$ or $1$ according to the measurement result should the detector clicks. Repeating above steps for sufficient times, Alice and Bob announce their $i,j$ and $m,n$ for each trial and only retain the raw key bits correspond to the cases that $\{i,j\}=\{m,n\}$ as their sifted key bits. In Ref.~\cite{chau15}, Chau proved that secret key bits can be generated when the error rate of sifted key bits is very high. It is remarkable  that this scheme can tolerate up to $50\%$ error rate provided that $L = 2^p$ for some integer $p\geqslant 2$.

 It is instructive to realize the Chau15 protocol experimentally. One challenge is that one must prepare and measure certain high-dimensional quantum states, namely, qudits, in the Chau15 protocol although preparing and measuring these special qudit states are less complicated that those for a general qudit state. Another challenge is that the two-way classical communication post-processing method reported in Ref.~\cite{chau15} is both rather complicated and of low yield. Here, we first present a new security proof of the Chau15 protocol based on standard one-way communication plus a simple secret key rate formula for any integer $L\geqslant 4$ rather than only for $L$ in the form $2^p$. Then, we report a proof-of-principle experimental demonstration of the Chau15 protocol. In this experiment, we encode the qudit based on quantum superposition of time-bin and measurement is performed with the help of a variable delay interferometer \cite{RRDPSexp2}.  As far as we know, this is the first experiment of the Chau15 protocol.

{\it Security proof of the Chau15 protocol for arbitrary $L\geqslant 4$ with one-way communication.} We first consider the case of an ideal single photon sourceand ideal detectors.
Since the protocol is permutationally symmetric, quantum de~Finetti
theorem~\cite{definetti} implies that we only need to consider the security under Eve's general collective attack in the form
\begin{equation}
\label{attack}
\begin{aligned}
&U_\text{Eve}\ket{i}\ket{E_{00}}=\sum^L_{j=0}c_{ij}\ket{j}\ket{E_{ij}} , \\
\end{aligned}
\end{equation}
where $\ket{E_{ij}}$ is the quantum state of Eve's ancilla, $\{\ket{E_{ij}}\}$ is a set of basis for Eve's ancilla, $\braket{E_{ij}}{E_{il}}=\delta_{jl}$. Without loss of generality, we assume $c_{ij}\geqslant 0$ and $\sum^N_{j=0}c_{ij}^2=1$.

Denote the probability that Bob obtains $\ket{\psi^\pm_{mn}}$ conditioned on the facts that Alice prepares her state as $\ket{\psi^\pm_{ij}}$ and Bob tried to project the state along $\ket{\psi^\pm_{mn}}$ by $p(m,n|i,j)$. In other words, $p(m,n|i,j)$ is the chance that a quantum state in Hilbert space spanned by $\{\ket{i},\ket{j}\}$ is transformed into a space spanned by $\{\ket{m},\ket{n}\}$.

In the Supplemental Material, we show that Eve's information on all sifted key bits is given by
\begin{equation}
\label{IAE2}
\begin{aligned}
I_{AE}\leqslant h_2(\frac{\sum_{i<j,m<n,m,n \ne i,j}p(m,n|i,j)}{(L-2)(L-3)\sum_{i<j}p(i,j|i,j)}) .
\end{aligned}
\end{equation}
For easy use by experimentalists, we define the mean counting rate $Q=\sum_{i<j}p(i,j|i,j)/\binom{L}{2}$ and $Q'=\sum_{i<j,m<n,m,n \ne i,j}p(m,n|i,j)/(\binom{L}{2}\binom{L-2}{2})$ (where $\binom{x}{y}$ is the bionomial coefficient), then we have

\begin{equation}
\label{IAE3}
\begin{aligned}
I_{AE}\leqslant h_2(\frac{Q}{2Q'}).
\end{aligned}
\end{equation}
Finally, the secret key rate per sifted key bit is given by $R=1-h_2(E)-I_{AE}$, where $E$ is the bit error rate of the sifted key bit.

We remark that our asymptotic bound of $I_{AE}$ holds even when $(i,j)$ is biased distributed. As long as the probability for Bob to measure any one of the ${(i,j)}$ pairs is non-zero, we could then estimate $p(m,n|i,j)$ to arbitrarily good precision given a sufficiently long key~\cite{Lo:EffBB84:2005}.
With the help of decoy states~\cite{Hwang:Decoy:2003,Wang:Decoy:2005,Lo:Decoy:2005}, this security proof can be adepted in real-life implementations with weak coherent sources.
And in the Supplementary Material, we extend our analysis to the case of finite key length with decoy states.

\begin{figure}[!h]\center
	\resizebox{8cm}{!}{\includegraphics{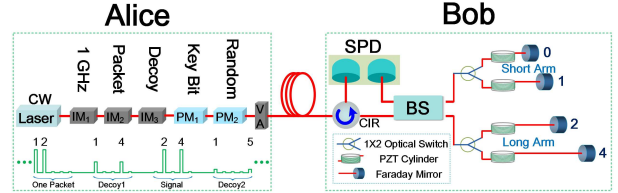}} \caption{Experimental setup for the Chau15 experiment. IM: intensity modulator; PM: phase modulator; VA: variable attenuator; CIR: circulator; OS: optical switch; BS: beam splitter; SPD: single photon detector.}\label{schematic}
\end{figure}

{\it Experimental setup and results.} We performed the experiment using the setup in Fig.~\ref{schematic}.
Alice consists of a continuous wave (CW) laser at 1550.12 nm, three intensity modulators (IM) and two phase modulators (PM). IM$_1$ modulates the CW light into a coherent pulse train with a temporal width of $96$~ps and a repetition rate of $1$~GHz. IM$_2$ chops this pulse train into packets of 5 time slots (5 ns), in which only two random pulses indexed by $i$ and $j$ ($i,j\in\{1,2,3,4,5\}$ and $i<j$) are allowed to pass. IM$_3$ is employed to implement the decoy states method \cite{Hwang:Decoy:2003,Wang:Decoy:2005,Lo:Decoy:2005}, in which each packet is randomly modulated into signal, decoy1, or decoy2 states. PM$_1$ encodes key bits by modulating phase $\{-\frac{\pi}{2},\frac{\pi}{2}\}$ on pulses for each packet, and PM$_2$ adds random global phase on each packet. Finally, the variable attenuator (VA) attenuates the average photon numbers per packet to the single photon level. The output quantum state prepared by Alice can be expressed as $\ket{\psi_{ij}}=(\ket{i}+e^{i\phi}\ket{j})/\sqrt{2}$, where $\phi\in\{0,\pi\}$ is the phase difference between the $i$th and $j$th pulses.

Bob mainly consists of a $1$~GHz, $1-4$~bit variable-delay Faraday-Michelson interferometer (FMI) and a double-channel single-photon detector (SPD). By setting the delay to $r$, the variable-delay FMI makes the $i$th pulse interfere with the $j$th pulse if $j=i+r$. Two channels of the SPD are connected to each of the output ports of the FMI, and which channel records a detection event depends on the phase difference $\phi$. Through recording a click event in the interference slot, Bob fulfills the quantum projection into $\ket{\psi^{\pm}_{ij}}$.

To realize the variable-delay interferometer, the same structure as in Ref.~\cite{RRDPSexp2} is employed here. Each of the two arms of this interferometer has two fiber delays, and the delay is chosen by a bidirectional NanoSpeed $1\times2$ optical switch (OS). The arm with the $\{0,1\}$ delays is named the short arm, and the arm with the $\{2,4\}$ delays is named the long arm. The chosen delays of the long and short arms are denoted as $x\in\{2,4\}$ and $y\in\{0,1\}$ respectively. Then, the delay of the interferometer is $r=x-y$. In total, the variable-delay FMI can achieve fast switch among $1-4$~ns delay values. The piezoelectric transducer (PZT) cylinder wrapped with $2$~m fiber was used to obtain high precision of each delay, and also to compensate the phase shift of the interferometer. The insertion loss (IL) of the interferometer (including the IL of the circulator) is about $2.0$~dB, and the values of each delay are almost the same. This variable-delay interferometer is polarization insensitive owning to Faraday mirrors and features an average extinction ratio of $23$~dB.

Photons from the variable-delay FMI were detected by the double-channel SPD, and finally recorded by a time-to-digital converter (TDC). Both channels of SPD are based on InGaAs/InP avalanche photodiodes and employ the sine-wave filtering method \cite{he2017sine}. The first channel features a detection efficiency of $22.1\%$ (down to $20.8\%$ if including $2\rightarrow3$ IL of CIR), a dark count rate of $1.5\times10^{-6}$ per gate, and after-pulse probability of $0.8\%$. The second channel features a detection efficiency of $20.9\%$, a dark count rate of $1.1\times10^{-6}$ per gate, and after-pulse probability of $1.1\%$. The TDC not only recorded signals but also set measurement time window. The value of the time window was set to $800$~ps during the experiment, and this setting reduced the average effective detection efficiency of two channels of the SPD to $20.4\%$, while the total dark count was just a little less than $2.6\times10^{-6}$ per gate.

\begin{table*}
	\centering
	\caption{Experimental results. The length of fiber ($l$), intensity ($Int.$) and probability ($P$) of one signal ($\mu$) and two decoy states ($\nu_1$ and $\nu_2$), the mean yield ($Q$ and $Q'$), error rate of the signal state ($E_\mu$), and the secret key rate per packet($R_{inf}$ for infinite packet number and $R_{f}$ for finite packet number). The last set of data over $50\ km$ fiber is obtained in high error rate case. The total number of packets sent from Alice is $N=3\times10^{11}$, the failure probability is set to be $10^{-10}$, and parameters are optimized.}	\label{results}
	\renewcommand\arraystretch{1.25}
	\setlength{\tabcolsep}{6pt}
	\begin{tabular}{|c|c|c|c|c|c|c|c|c|}
		\hline
		$l(km)$ &\multicolumn{2}{c|}{$Int.(ph/packet)$} & $P(\%)$ & $Q$ & $Q'$ & $E_\mu$(\%) & $R_{inf}$ & $R_f$\\
		\hline
		\multirow{3}{*}{$50$} & $\mu$ & $0.66$ & $97.81$ & $4.36\times10^{-3}$ & $1.10\times10^{-5}$ &\multirow{3}{*}{$1.83$} & \multirow{3}{*}{$1.45\times10^{-3}$} & \multirow{3}{*}{$1.39\times10^{-3}$}\\
		& $\nu_1$ & $0.05$ & $1.40$ & $3.33\times10^{-4}$ & $3.23\times10^{-6}$ & & & \\
		& $\nu_2$ & $0.0016$ & $0.79$ & $1.34\times10^{-5}$ & $2.61\times10^{-6}$ & & & \\
		\hline
		
		\multirow{3}{*}{$100$} & $\mu$ & $0.62$ & $94.64$ & $4.05\times10^{-4}$ & $3.36\times10^{-6}$ &\multirow{3}{*}{$2.16$} & \multirow{3}{*}{$1.20\times10^{-4}$} & \multirow{3}{*}{$1.06\times10^{-4}$}\\
		& $\nu_1$ & $0.10$ & $3.36$ & $6.75\times10^{-5}$ & $2.71\times10^{-6}$ & & & \\
		& $\nu_2$ & $0.0015$ & $2.00$ & $3.58\times10^{-6}$ & $2.60\times10^{-6}$ & & & \\
		\hline
		
		\multirow{3}{*}{$130$} & $\mu$ & $0.57$ & $87.74$ & $9.04\times10^{-5}$ & $2.76\times10^{-6}$ &\multirow{3}{*}{$3.21$} & \multirow{3}{*}{$1.73\times10^{-5}$} & \multirow{3}{*}{$1.28\times10^{-5}$}\\
		& $\nu_1$ & $0.14$ & $7.52$ & $2.42\times10^{-5}$ & $2.63\times10^{-6}$ & & & \\
		& $\nu_2$ & $0.0014$ & $4.74$ & $2.81\times10^{-6}$ & $2.60\times10^{-6}$ & & & \\
		\hline
		
		\multirow{3}{*}{$150$} & $\mu$ & $0.50$ & $36.12$ & $3.31\times10^{-5}$ & $2.64\times10^{-6}$ &\multirow{3}{*}{$5.68$} & \multirow{3}{*}{$4.32\times10^{-7}$} & \multirow{3}{*}{$--$}\\
		& $\nu_1$ & $0.14$ & $37.76$ & $1.11\times10^{-5}$ & $2.61\times10^{-6}$ & & & \\
		& $\nu_2$ & $0.0012$ & $26.12$ & $2.67\times10^{-6}$ & $2.60\times10^{-6}$ & & & \\
		\hline
		
		\multirow{3}{*}{$50$} & $\mu$ & $0.07$ & $84.45$ & $4.65\times10^{-4}$ & $3.47\times10^{-6}$ &\multirow{3}{*}{$20.32$} & \multirow{3}{*}{$2.40\times10^{-5}$} & \multirow{3}{*}{$1.70\times10^{-5}$}\\
		& $\nu_1$ & $0.035$ & $10.40$ & $2.34\times10^{-4}$ & $3.03\times10^{-6}$ & & & \\
		& $\nu_2$ & $0.0002$ & $5.15$ & $3.73\times10^{-6}$ & $2.60\times10^{-6}$ & & & \\
		\hline
		
		\end{tabular}
		\end{table*}
		
Based on the experimental parameters listed above, the performance of Chau15 system was estimated and all parameters were optimized by maximizing the secret key rate. Setting the intensity and probability of one signal ($\mu$) and two decoy states ($\nu_1$ and $\nu_2$) close to the optimal ones, we measured the mean yield $Q$ and $Q'$, and error rate of the signal state $E_{\mu}$ at four fiber lengths: $50$, $100$, $130$ and $150\ km$. When Alice prepares laser pulse at time slots $i$ and $j$, Bob gets the mean yield $Q=\sum_{i<j}p(i,j|i,j)/\binom{5}{2}$ if he also gets interference output between these two time slots, and Bob gets the mean yield  $Q'=\sum_{i<j,m<n}p(m,n|i,j)/(\binom{5}{2}\binom{3}{2})$ if he obtains interference output between time slots $m$ and $n$ ($m,n\neq i,j$). The experimental results are listed in Table~\ref{results}. The last set of data over $50\ km$ length fiber is obtained in high error rate case, which may happen in the ultrahigh speed case or harsher environment. In the experiment, we intentionally distorted the modulating signal on $PM_1$ to get error rate over $20\%$.
		
Secret keys can still be extracted at a transmission distance of $150\ km$ in the asymptotic case ($R_{inf}$ when Alice sends infinite packets), which is comparable with the commonly used BB84 protocol. The secret key rate per packet is at $10^{-3}$ level over $50\ km$ length fiber, which is lower than BB84 protocol, but if the biased-basis method was employed in Chau15 protocol, its secret key rate is also comparable with BB84 in the asymptotic case. When the error rate exceeds $20\%$, Chau15 protocol can still get secret key rate at $10^{-5}$ level over $50\ km$ length fiber. Therefore, Chau15 protocol can outperform BB84, especially in the high error rate case. And, Chau15 protocol can also outperform RRDPS. The maximum transmission distance of the $L=5$ RRDPS experiment system is less than $50\ km$ using superconducting SPDs \cite{RRDPSexp1}, while in our $L=5$ Chau15 system, the transmission distance could reach $150\ km$ with InGaAs/InP SPD. The tolerant error rate of the $L=65$ RRDPS experiment system with the weak coherent source is less than $17\%$ \cite{RRDPSexp2}, while the value can exceed $20\%$ in Chau15 system even $L=5$. These results verified that Chau15 protocol has good comprehensive performances on maximum transmission distance, secret key rate and tolerant error rate with small $L$.
		
Nevertheless, the secret key rate per second of our proof-of-principle experimental realization is limited by the optical switch, whose switching speed in the setup is about $100\ ns$. This limitation can be overcame by employing the passive scheme based on $1\times(L-1)$	BS (just like \cite{RRDPSexp1}), or slow basis choice method \cite{sasaki2016quantum}, or the development of optical switch techniques in the near future. 	
		
\begin{figure}[!h]\center
	\resizebox{8cm}{!}{\includegraphics{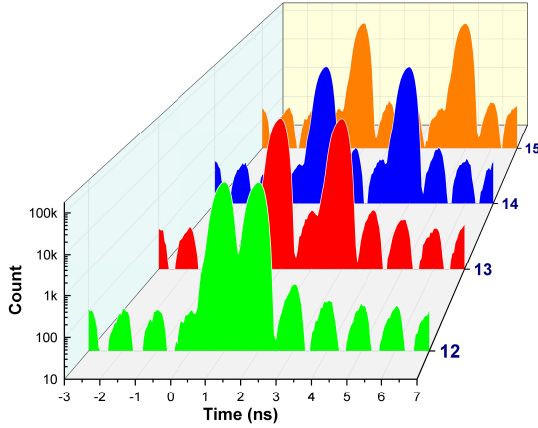}} \caption{Typical outputs of Alice detected by Bob's SPD (without interferometer). Here four kinds of typical outputs are given, $\mathbf{12}$ means $i=1,j=2$, $\cdots$, $\mathbf{15}$ means $i=1,j=5$. }\label{schematic2}
\end{figure}

Another difficulty to overcome is to obtain relatively small yield $Q'$, which plays an important role to estimate Eve's information (see Eq.(\ref{IAE3})). The smaller $Q'$ Bob measures, the more key rate Alice and Bob can share. Under ideal conditions, the yield $Q'$ should be equal to the dark count of the SPD. However, the experimental results show that the yield $Q'$ is larger than the dark count rate of the SPD, especially at short transmission distance. To offer an intuitive impression, we define the count ratio $C_m$ as the count at time slot $i$ to the count at time slot $m\neq i,j$ for the output packets from Alice. $Q'$ can be evaluated by the mean value of $C_m$. The sources contributing to relatively large $Q'$ mainly include the limited extinction ratio of $IM_2$, the dark count, after pulses and time jitter of the SPD. Therefore, Bob's SPD (without interferometer) is employed to directly measure the outputs of Alice. Four kinds of typical outputs with  $i=1$ ($\mathbf{12}$, $\mathbf{13}$, $\mathbf{14}$, and $\mathbf{15}$)  are shown in Fig.~\ref{schematic2}, the time window of TDC is not set for this measurement. The worst case corresponds to the outputs of $j=i+2$. Taking $\mathbf{13}$ output of Alice for example, the count ratio $C_2$ is only $280$.

{\it Conclusion.} In summary, we developed a security proof for the Chau15 protocol with one-way post-processing, which facilitates the secret key generation in real-life situation. In our experiment, the qubit-like qudits are prepared by manipulating time-bin of photon and measured by a variable delay interferometer. Our demonstration exhibits the fine feasibility and high error-rate tolerance of the novel Chau15 protocol and sheds light on QKD experiments with high-dimensionality.

{\it Acknowledgements.} S.W. and Z.-Q.Y. contributed equally to this work. The USTC team is supported by the National Natural Science Foundation of China (Grant Nos. 61622506, 61475148, 61575183, 61627820, 61675189), the National Key Research And Development Program of China (Grant Nos. 2016YFA0302600, 2016YFA0301702), the "Strategic Priority Research Program(B)" of the Chinese Academy of Sciences (Grant No. XDB01030100). H.F.C. is supported by the Research Grant Council of the HKSAR Government (Grant No. 17304716).

\bibliographystyle{apsrev4-1}

\bibliography{Biblisource}

\begin{thebibliography}{29}%
\makeatletter
\providecommand \@ifxundefined [1]{%
 \@ifx{#1\undefined}
}%
\providecommand \@ifnum [1]{%
 \ifnum #1\expandafter \@firstoftwo
 \else \expandafter \@secondoftwo
 \fi
}%
\providecommand \@ifx [1]{%
 \ifx #1\expandafter \@firstoftwo
 \else \expandafter \@secondoftwo
 \fi
}%
\providecommand \natexlab [1]{#1}%
\providecommand \enquote  [1]{``#1''}%
\providecommand \bibnamefont  [1]{#1}%
\providecommand \bibfnamefont [1]{#1}%
\providecommand \citenamefont [1]{#1}%
\providecommand \href@noop [0]{\@secondoftwo}%
\providecommand \href [0]{\begingroup \@sanitize@url \@href}%
\providecommand \@href[1]{\@@startlink{#1}\@@href}%
\providecommand \@@href[1]{\endgroup#1\@@endlink}%
\providecommand \@sanitize@url [0]{\catcode `\\12\catcode `\$12\catcode
  `\&12\catcode `\#12\catcode `\^12\catcode `\_12\catcode `\%12\relax}%
\providecommand \@@startlink[1]{}%
\providecommand \@@endlink[0]{}%
\providecommand \url  [0]{\begingroup\@sanitize@url \@url }%
\providecommand \@url [1]{\endgroup\@href {#1}{\urlprefix }}%
\providecommand \urlprefix  [0]{URL }%
\providecommand \Eprint [0]{\href }%
\providecommand \doibase [0]{http://dx.doi.org/}%
\providecommand \selectlanguage [0]{\@gobble}%
\providecommand \bibinfo  [0]{\@secondoftwo}%
\providecommand \bibfield  [0]{\@secondoftwo}%
\providecommand \translation [1]{[#1]}%
\providecommand \BibitemOpen [0]{}%
\providecommand \bibitemStop [0]{}%
\providecommand \bibitemNoStop [0]{.\EOS\space}%
\providecommand \EOS [0]{\spacefactor3000\relax}%
\providecommand \BibitemShut  [1]{\csname bibitem#1\endcsname}%
\let\auto@bib@innerbib\@empty
\bibitem [{\citenamefont {Chau}(2015)}]{chau15}%
  \BibitemOpen
  \bibfield  {author} {\bibinfo {author} {\bibfnamefont {H.~F.}\ \bibnamefont
  {Chau}},\ }\href {\doibase 10.1103/PhysRevA.92.062324} {\bibfield  {journal}
  {\bibinfo  {journal} {Phys. Rev. A}\ }\textbf {\bibinfo {volume} {92}},\
  \bibinfo {pages} {062324} (\bibinfo {year} {2015})}\BibitemShut {NoStop}%
\bibitem [{\citenamefont {Bennett}\ and\ \citenamefont
  {Brassard}(1984)}]{Bennett:BB84:1984}%
  \BibitemOpen
  \bibfield  {author} {\bibinfo {author} {\bibfnamefont {C.~H.}\ \bibnamefont
  {Bennett}}\ and\ \bibinfo {author} {\bibfnamefont {G.}~\bibnamefont
  {Brassard}},\ }in\ \href@noop {} {\emph {\bibinfo {booktitle} {Proceedings of
  the IEEE International Conference on Computers, Systems and Signal
  Processing}}}\ (\bibinfo  {publisher} {IEEE Press},\ \bibinfo {address} {New
  York},\ \bibinfo {year} {1984})\ pp.\ \bibinfo {pages} {175--179}\BibitemShut
  {NoStop}%
\bibitem [{\citenamefont {Ekert}(1991)}]{Ekert:QKD:1991}%
  \BibitemOpen
  \bibfield  {author} {\bibinfo {author} {\bibfnamefont {A.~K.}\ \bibnamefont
  {Ekert}},\ }\href {\doibase 10.1103/PhysRevLett.67.661} {\bibfield  {journal}
  {\bibinfo  {journal} {Phys. Rev. Lett.}\ }\textbf {\bibinfo {volume} {67}},\
  \bibinfo {pages} {661} (\bibinfo {year} {1991})}\BibitemShut {NoStop}%
\bibitem [{\citenamefont {Hwang}(2003)}]{Hwang:Decoy:2003}%
  \BibitemOpen
  \bibfield  {author} {\bibinfo {author} {\bibfnamefont {W.-Y.}\ \bibnamefont
  {Hwang}},\ }\href {\doibase 10.1103/PhysRevLett.91.057901} {\bibfield
  {journal} {\bibinfo  {journal} {Phys. Rev. Lett.}\ }\textbf {\bibinfo
  {volume} {91}},\ \bibinfo {pages} {057901} (\bibinfo {year}
  {2003})}\BibitemShut {NoStop}%
\bibitem [{\citenamefont {Wang}(2005)}]{Wang:Decoy:2005}%
  \BibitemOpen
  \bibfield  {author} {\bibinfo {author} {\bibfnamefont {X.-B.}\ \bibnamefont
  {Wang}},\ }\href@noop {} {\bibfield  {journal} {\bibinfo  {journal}
  {Phys.~Rev.~Lett.~}\ }\textbf {\bibinfo {volume} {94}},\ \bibinfo {pages}
  {230503} (\bibinfo {year} {2005})}\BibitemShut {NoStop}%
\bibitem [{\citenamefont {Lo}\ \emph {et~al.}(2005{\natexlab{a}})\citenamefont
  {Lo}, \citenamefont {Ma},\ and\ \citenamefont {Chen}}]{Lo:Decoy:2005}%
  \BibitemOpen
  \bibfield  {author} {\bibinfo {author} {\bibfnamefont {H.-K.}\ \bibnamefont
  {Lo}}, \bibinfo {author} {\bibfnamefont {X.}~\bibnamefont {Ma}}, \ and\
  \bibinfo {author} {\bibfnamefont {K.}~\bibnamefont {Chen}},\ }\href@noop {}
  {\bibfield  {journal} {\bibinfo  {journal} {Phys.~Rev.~Lett.~}\ }\textbf
  {\bibinfo {volume} {94}},\ \bibinfo {pages} {230504} (\bibinfo {year}
  {2005}{\natexlab{a}})}\BibitemShut {NoStop}%
\bibitem [{\citenamefont {Zhao}\ \emph
  {et~al.}(2006{\natexlab{a}})\citenamefont {Zhao}, \citenamefont {Qi},
  \citenamefont {Ma}, \citenamefont {Lo},\ and\ \citenamefont
  {Qian}}]{Zhao:DecoyExp:2006}%
  \BibitemOpen
  \bibfield  {author} {\bibinfo {author} {\bibfnamefont {Y.}~\bibnamefont
  {Zhao}}, \bibinfo {author} {\bibfnamefont {B.}~\bibnamefont {Qi}}, \bibinfo
  {author} {\bibfnamefont {X.}~\bibnamefont {Ma}}, \bibinfo {author}
  {\bibfnamefont {H.-K.}\ \bibnamefont {Lo}}, \ and\ \bibinfo {author}
  {\bibfnamefont {L.}~\bibnamefont {Qian}},\ }\href@noop {} {\bibfield
  {journal} {\bibinfo  {journal} {Phys.~Rev.~Lett.~}\ }\textbf {\bibinfo
  {volume} {96}},\ \bibinfo {pages} {070502} (\bibinfo {year}
  {2006}{\natexlab{a}})}\BibitemShut {NoStop}%
\bibitem [{\citenamefont {Rosenberg}\ \emph {et~al.}(2007)\citenamefont
  {Rosenberg}, \citenamefont {Harrington}, \citenamefont {Rice}, \citenamefont
  {Hiskett}, \citenamefont {Peterson}, \citenamefont {Hughes}, \citenamefont
  {Lita}, \citenamefont {Nam},\ and\ \citenamefont
  {Nordholt}}]{Rosenberg:ExpDecoy:2007}%
  \BibitemOpen
  \bibfield  {author} {\bibinfo {author} {\bibfnamefont {D.}~\bibnamefont
  {Rosenberg}}, \bibinfo {author} {\bibfnamefont {J.~W.}\ \bibnamefont
  {Harrington}}, \bibinfo {author} {\bibfnamefont {P.~R.}\ \bibnamefont
  {Rice}}, \bibinfo {author} {\bibfnamefont {P.~A.}\ \bibnamefont {Hiskett}},
  \bibinfo {author} {\bibfnamefont {C.~G.}\ \bibnamefont {Peterson}}, \bibinfo
  {author} {\bibfnamefont {R.~J.}\ \bibnamefont {Hughes}}, \bibinfo {author}
  {\bibfnamefont {A.~E.}\ \bibnamefont {Lita}}, \bibinfo {author}
  {\bibfnamefont {S.~W.}\ \bibnamefont {Nam}}, \ and\ \bibinfo {author}
  {\bibfnamefont {J.~E.}\ \bibnamefont {Nordholt}},\ }\href {\doibase
  10.1103/PhysRevLett.98.010503} {\bibfield  {journal} {\bibinfo  {journal}
  {Phys. Rev. Lett.}\ }\textbf {\bibinfo {volume} {98}},\ \bibinfo {pages}
  {010503} (\bibinfo {year} {2007})}\BibitemShut {NoStop}%
\bibitem [{\citenamefont {Schmitt-Manderbach}\ \emph
  {et~al.}(2007)\citenamefont {Schmitt-Manderbach}, \citenamefont {Weier},
  \citenamefont {F\"urst}, \citenamefont {Ursin}, \citenamefont {Tiefenbacher},
  \citenamefont {Scheidl}, \citenamefont {Perdigues}, \citenamefont {Sodnik},
  \citenamefont {Kurtsiefer}, \citenamefont {Rarity}, \citenamefont
  {Zeilinger},\ and\ \citenamefont {Weinfurter}}]{Zeilinger:ExpDecoy:2007}%
  \BibitemOpen
  \bibfield  {author} {\bibinfo {author} {\bibfnamefont {T.}~\bibnamefont
  {Schmitt-Manderbach}}, \bibinfo {author} {\bibfnamefont {H.}~\bibnamefont
  {Weier}}, \bibinfo {author} {\bibfnamefont {M.}~\bibnamefont {F\"urst}},
  \bibinfo {author} {\bibfnamefont {R.}~\bibnamefont {Ursin}}, \bibinfo
  {author} {\bibfnamefont {F.}~\bibnamefont {Tiefenbacher}}, \bibinfo {author}
  {\bibfnamefont {T.}~\bibnamefont {Scheidl}}, \bibinfo {author} {\bibfnamefont
  {J.}~\bibnamefont {Perdigues}}, \bibinfo {author} {\bibfnamefont
  {Z.}~\bibnamefont {Sodnik}}, \bibinfo {author} {\bibfnamefont
  {C.}~\bibnamefont {Kurtsiefer}}, \bibinfo {author} {\bibfnamefont {J.~G.}\
  \bibnamefont {Rarity}}, \bibinfo {author} {\bibfnamefont {A.}~\bibnamefont
  {Zeilinger}}, \ and\ \bibinfo {author} {\bibfnamefont {H.}~\bibnamefont
  {Weinfurter}},\ }\href {\doibase 10.1103/PhysRevLett.98.010504} {\bibfield
  {journal} {\bibinfo  {journal} {Phys. Rev. Lett.}\ }\textbf {\bibinfo
  {volume} {98}},\ \bibinfo {pages} {010504} (\bibinfo {year}
  {2007})}\BibitemShut {NoStop}%
\bibitem [{\citenamefont {Peng}\ \emph {et~al.}(2007)\citenamefont {Peng},
  \citenamefont {Zhang}, \citenamefont {Yang}, \citenamefont {Gao},
  \citenamefont {Ma}, \citenamefont {Yin}, \citenamefont {Zeng}, \citenamefont
  {Yang}, \citenamefont {Wang},\ and\ \citenamefont
  {Pan}}]{Peng:ExpDecoy:2007}%
  \BibitemOpen
  \bibfield  {author} {\bibinfo {author} {\bibfnamefont {C.-Z.}\ \bibnamefont
  {Peng}}, \bibinfo {author} {\bibfnamefont {J.}~\bibnamefont {Zhang}},
  \bibinfo {author} {\bibfnamefont {D.}~\bibnamefont {Yang}}, \bibinfo {author}
  {\bibfnamefont {W.-B.}\ \bibnamefont {Gao}}, \bibinfo {author} {\bibfnamefont
  {H.-X.}\ \bibnamefont {Ma}}, \bibinfo {author} {\bibfnamefont
  {H.}~\bibnamefont {Yin}}, \bibinfo {author} {\bibfnamefont {H.-P.}\
  \bibnamefont {Zeng}}, \bibinfo {author} {\bibfnamefont {T.}~\bibnamefont
  {Yang}}, \bibinfo {author} {\bibfnamefont {X.-B.}\ \bibnamefont {Wang}}, \
  and\ \bibinfo {author} {\bibfnamefont {J.-W.}\ \bibnamefont {Pan}},\ }\href
  {\doibase 10.1103/PhysRevLett.98.010505} {\bibfield  {journal} {\bibinfo
  {journal} {Phys. Rev. Lett.}\ }\textbf {\bibinfo {volume} {98}},\ \bibinfo
  {pages} {010505} (\bibinfo {year} {2007})}\BibitemShut {NoStop}%
\bibitem [{\citenamefont {Zhao}\ \emph
  {et~al.}(2006{\natexlab{b}})\citenamefont {Zhao}, \citenamefont {Qi},
  \citenamefont {Ma}, \citenamefont {Lo},\ and\ \citenamefont
  {Qian}}]{Zhao:Decoy60km:2006}%
  \BibitemOpen
  \bibfield  {author} {\bibinfo {author} {\bibfnamefont {Y.}~\bibnamefont
  {Zhao}}, \bibinfo {author} {\bibfnamefont {B.}~\bibnamefont {Qi}}, \bibinfo
  {author} {\bibfnamefont {X.}~\bibnamefont {Ma}}, \bibinfo {author}
  {\bibfnamefont {H.-K.}\ \bibnamefont {Lo}}, \ and\ \bibinfo {author}
  {\bibfnamefont {L.}~\bibnamefont {Qian}},\ }in\ \href@noop {} {\emph
  {\bibinfo {booktitle} {Proc.~of IEEE ISIT}}}\ (\bibinfo  {publisher} {IEEE},\
  \bibinfo {year} {2006})\ p.\ \bibinfo {pages} {2094}\BibitemShut {NoStop}%
\bibitem [{\citenamefont {Yuan}\ \emph {et~al.}(2007)\citenamefont {Yuan},
  \citenamefont {Sharpe},\ and\ \citenamefont {Shields}}]{Yuan:ExpDecoy2007}%
  \BibitemOpen
  \bibfield  {author} {\bibinfo {author} {\bibfnamefont {Z.~L.}\ \bibnamefont
  {Yuan}}, \bibinfo {author} {\bibfnamefont {A.~W.}\ \bibnamefont {Sharpe}}, \
  and\ \bibinfo {author} {\bibfnamefont {A.~J.}\ \bibnamefont {Shields}},\
  }\href@noop {} {\bibfield  {journal} {\bibinfo  {journal}
  {Appl.~Phys.~Lett.}\ }\textbf {\bibinfo {volume} {90}},\ \bibinfo {pages}
  {011118} (\bibinfo {year} {2007})}\BibitemShut {NoStop}%
\bibitem [{\citenamefont {Frolich}\ \emph {et~al.}(2013)\citenamefont
  {Frolich}, \citenamefont {Dynes}, \citenamefont {Lucamarini}, \citenamefont
  {Sharpe}, \citenamefont {Yuan},\ and\ \citenamefont
  {Shields}}]{Frolich:QKDnet:2013}%
  \BibitemOpen
  \bibfield  {author} {\bibinfo {author} {\bibfnamefont {B.}~\bibnamefont
  {Frolich}}, \bibinfo {author} {\bibfnamefont {J.~F.}\ \bibnamefont {Dynes}},
  \bibinfo {author} {\bibfnamefont {M.}~\bibnamefont {Lucamarini}}, \bibinfo
  {author} {\bibfnamefont {A.~W.}\ \bibnamefont {Sharpe}}, \bibinfo {author}
  {\bibfnamefont {Z.}~\bibnamefont {Yuan}}, \ and\ \bibinfo {author}
  {\bibfnamefont {A.~J.}\ \bibnamefont {Shields}},\ }\href {\doibase
  10.1038/nature12493} {\bibfield  {journal} {\bibinfo  {journal} {Nature}\
  }\textbf {\bibinfo {volume} {501}},\ \bibinfo {pages} {69} (\bibinfo {year}
  {2013})}\BibitemShut {NoStop}%
\bibitem [{\citenamefont {Lo}\ \emph {et~al.}(2012)\citenamefont {Lo},
  \citenamefont {Curty},\ and\ \citenamefont {Qi}}]{Lo:MDIQKD:2012}%
  \BibitemOpen
  \bibfield  {author} {\bibinfo {author} {\bibfnamefont {H.-K.}\ \bibnamefont
  {Lo}}, \bibinfo {author} {\bibfnamefont {M.}~\bibnamefont {Curty}}, \ and\
  \bibinfo {author} {\bibfnamefont {B.}~\bibnamefont {Qi}},\ }\href {\doibase
  10.1103/PhysRevLett.108.130503} {\bibfield  {journal} {\bibinfo  {journal}
  {Phys. Rev. Lett.}\ }\textbf {\bibinfo {volume} {108}},\ \bibinfo {pages}
  {130503} (\bibinfo {year} {2012})}\BibitemShut {NoStop}%
\bibitem [{\citenamefont {Braunstein}\ and\ \citenamefont
  {Pirandola}(2012)}]{Braunstein:MIQKD:2012}%
  \BibitemOpen
  \bibfield  {author} {\bibinfo {author} {\bibfnamefont {S.~L.}\ \bibnamefont
  {Braunstein}}\ and\ \bibinfo {author} {\bibfnamefont {S.}~\bibnamefont
  {Pirandola}},\ }\href {\doibase 10.1103/PhysRevLett.108.130502} {\bibfield
  {journal} {\bibinfo  {journal} {Phys. Rev. Lett.}\ }\textbf {\bibinfo
  {volume} {108}},\ \bibinfo {pages} {130502} (\bibinfo {year}
  {2012})}\BibitemShut {NoStop}%
\bibitem [{\citenamefont {Sasaki}\ \emph {et~al.}(2014)\citenamefont {Sasaki},
  \citenamefont {Yamamoto},\ and\ \citenamefont
  {Koashi}}]{sasaki2014practical}%
  \BibitemOpen
  \bibfield  {author} {\bibinfo {author} {\bibfnamefont {T.}~\bibnamefont
  {Sasaki}}, \bibinfo {author} {\bibfnamefont {Y.}~\bibnamefont {Yamamoto}}, \
  and\ \bibinfo {author} {\bibfnamefont {M.}~\bibnamefont {Koashi}},\
  }\href@noop {} {\bibfield  {journal} {\bibinfo  {journal} {Nature}\ }\textbf
  {\bibinfo {volume} {509}},\ \bibinfo {pages} {475} (\bibinfo {year}
  {2014})}\BibitemShut {NoStop}%
\bibitem [{\citenamefont {Rubenok}\ \emph {et~al.}(2013)\citenamefont
  {Rubenok}, \citenamefont {Slater}, \citenamefont {Chan}, \citenamefont
  {Lucio-Martinez},\ and\ \citenamefont {Tittel}}]{Rubenok:MIQKDexp:2013}%
  \BibitemOpen
  \bibfield  {author} {\bibinfo {author} {\bibfnamefont {A.}~\bibnamefont
  {Rubenok}}, \bibinfo {author} {\bibfnamefont {J.~A.}\ \bibnamefont {Slater}},
  \bibinfo {author} {\bibfnamefont {P.}~\bibnamefont {Chan}}, \bibinfo {author}
  {\bibfnamefont {I.}~\bibnamefont {Lucio-Martinez}}, \ and\ \bibinfo {author}
  {\bibfnamefont {W.}~\bibnamefont {Tittel}},\ }\href {\doibase
  10.1103/PhysRevLett.111.130501} {\bibfield  {journal} {\bibinfo  {journal}
  {Phys. Rev. Lett.}\ }\textbf {\bibinfo {volume} {111}},\ \bibinfo {pages}
  {130501} (\bibinfo {year} {2013})}\BibitemShut {NoStop}%
\bibitem [{\citenamefont {Liu}\ \emph {et~al.}(2013)\citenamefont {Liu},
  \citenamefont {Chen}, \citenamefont {Wang}, \citenamefont {Liang},
  \citenamefont {Shentu}, \citenamefont {Wang}, \citenamefont {Cui},
  \citenamefont {Yin}, \citenamefont {Liu}, \citenamefont {Li}, \citenamefont
  {Ma}, \citenamefont {Pelc}, \citenamefont {Fejer}, \citenamefont {Peng},
  \citenamefont {Zhang},\ and\ \citenamefont {Pan}}]{liu:MIQKDexp:2013}%
  \BibitemOpen
  \bibfield  {author} {\bibinfo {author} {\bibfnamefont {Y.}~\bibnamefont
  {Liu}}, \bibinfo {author} {\bibfnamefont {T.-Y.}\ \bibnamefont {Chen}},
  \bibinfo {author} {\bibfnamefont {L.-J.}\ \bibnamefont {Wang}}, \bibinfo
  {author} {\bibfnamefont {H.}~\bibnamefont {Liang}}, \bibinfo {author}
  {\bibfnamefont {G.-L.}\ \bibnamefont {Shentu}}, \bibinfo {author}
  {\bibfnamefont {J.}~\bibnamefont {Wang}}, \bibinfo {author} {\bibfnamefont
  {K.}~\bibnamefont {Cui}}, \bibinfo {author} {\bibfnamefont {H.-L.}\
  \bibnamefont {Yin}}, \bibinfo {author} {\bibfnamefont {N.-L.}\ \bibnamefont
  {Liu}}, \bibinfo {author} {\bibfnamefont {L.}~\bibnamefont {Li}}, \bibinfo
  {author} {\bibfnamefont {X.}~\bibnamefont {Ma}}, \bibinfo {author}
  {\bibfnamefont {J.~S.}\ \bibnamefont {Pelc}}, \bibinfo {author}
  {\bibfnamefont {M.~M.}\ \bibnamefont {Fejer}}, \bibinfo {author}
  {\bibfnamefont {C.-Z.}\ \bibnamefont {Peng}}, \bibinfo {author}
  {\bibfnamefont {Q.}~\bibnamefont {Zhang}}, \ and\ \bibinfo {author}
  {\bibfnamefont {J.-W.}\ \bibnamefont {Pan}},\ }\href {\doibase
  10.1103/PhysRevLett.111.130502} {\bibfield  {journal} {\bibinfo  {journal}
  {Phys. Rev. Lett.}\ }\textbf {\bibinfo {volume} {111}},\ \bibinfo {pages}
  {130502} (\bibinfo {year} {2013})}\BibitemShut {NoStop}%
\bibitem [{\citenamefont {Ferreira~da Silva}\ \emph {et~al.}(2013)\citenamefont
  {Ferreira~da Silva}, \citenamefont {Vitoreti}, \citenamefont {Xavier},
  \citenamefont {do~Amaral}, \citenamefont {Tempor\~ao},\ and\ \citenamefont
  {von~der Weid}}]{daSilva:MIQKD:2013}%
  \BibitemOpen
  \bibfield  {author} {\bibinfo {author} {\bibfnamefont {T.}~\bibnamefont
  {Ferreira~da Silva}}, \bibinfo {author} {\bibfnamefont {D.}~\bibnamefont
  {Vitoreti}}, \bibinfo {author} {\bibfnamefont {G.~B.}\ \bibnamefont
  {Xavier}}, \bibinfo {author} {\bibfnamefont {G.~C.}\ \bibnamefont
  {do~Amaral}}, \bibinfo {author} {\bibfnamefont {G.~P.}\ \bibnamefont
  {Tempor\~ao}}, \ and\ \bibinfo {author} {\bibfnamefont {J.~P.}\ \bibnamefont
  {von~der Weid}},\ }\href {\doibase 10.1103/PhysRevA.88.052303} {\bibfield
  {journal} {\bibinfo  {journal} {Phys. Rev. A}\ }\textbf {\bibinfo {volume}
  {88}},\ \bibinfo {pages} {052303} (\bibinfo {year} {2013})}\BibitemShut
  {NoStop}%
\bibitem [{\citenamefont {Tang}\ \emph {et~al.}(2014)\citenamefont {Tang},
  \citenamefont {Liao}, \citenamefont {Xu}, \citenamefont {Qi}, \citenamefont
  {Qian},\ and\ \citenamefont {Lo}}]{Tang:MDIQKDexp:2013}%
  \BibitemOpen
  \bibfield  {author} {\bibinfo {author} {\bibfnamefont {Z.}~\bibnamefont
  {Tang}}, \bibinfo {author} {\bibfnamefont {Z.}~\bibnamefont {Liao}}, \bibinfo
  {author} {\bibfnamefont {F.}~\bibnamefont {Xu}}, \bibinfo {author}
  {\bibfnamefont {B.}~\bibnamefont {Qi}}, \bibinfo {author} {\bibfnamefont
  {L.}~\bibnamefont {Qian}}, \ and\ \bibinfo {author} {\bibfnamefont {H.-K.}\
  \bibnamefont {Lo}},\ }\href {\doibase 10.1103/PhysRevLett.112.190503}
  {\bibfield  {journal} {\bibinfo  {journal} {Phys. Rev. Lett.}\ }\textbf
  {\bibinfo {volume} {112}},\ \bibinfo {pages} {190503} (\bibinfo {year}
  {2014})}\BibitemShut {NoStop}%
\bibitem [{\citenamefont {Wang}\ \emph {et~al.}(2015)\citenamefont {Wang},
  \citenamefont {Song}, \citenamefont {Yin}, \citenamefont {Wang},
  \citenamefont {Chen}, \citenamefont {Zhang}, \citenamefont {Guo},\ and\
  \citenamefont {Han}}]{RFIMDIQKDexp}%
  \BibitemOpen
  \bibfield  {author} {\bibinfo {author} {\bibfnamefont {C.}~\bibnamefont
  {Wang}}, \bibinfo {author} {\bibfnamefont {X.-T.}\ \bibnamefont {Song}},
  \bibinfo {author} {\bibfnamefont {Z.-Q.}\ \bibnamefont {Yin}}, \bibinfo
  {author} {\bibfnamefont {S.}~\bibnamefont {Wang}}, \bibinfo {author}
  {\bibfnamefont {W.}~\bibnamefont {Chen}}, \bibinfo {author} {\bibfnamefont
  {C.-M.}\ \bibnamefont {Zhang}}, \bibinfo {author} {\bibfnamefont {G.-C.}\
  \bibnamefont {Guo}}, \ and\ \bibinfo {author} {\bibfnamefont {Z.-F.}\
  \bibnamefont {Han}},\ }\href {\doibase 10.1103/PhysRevLett.115.160502}
  {\bibfield  {journal} {\bibinfo  {journal} {Phys. Rev. Lett.}\ }\textbf
  {\bibinfo {volume} {115}},\ \bibinfo {pages} {160502} (\bibinfo {year}
  {2015})}\BibitemShut {NoStop}%
\bibitem [{\citenamefont {Guan}\ \emph {et~al.}(2015)\citenamefont {Guan},
  \citenamefont {Cao}, \citenamefont {Liu}, \citenamefont {Shen-Tu},
  \citenamefont {Pelc}, \citenamefont {Fejer}, \citenamefont {Peng},
  \citenamefont {Ma}, \citenamefont {Zhang},\ and\ \citenamefont
  {Pan}}]{PassiveRRDPS}%
  \BibitemOpen
  \bibfield  {author} {\bibinfo {author} {\bibfnamefont {J.-Y.}\ \bibnamefont
  {Guan}}, \bibinfo {author} {\bibfnamefont {Z.}~\bibnamefont {Cao}}, \bibinfo
  {author} {\bibfnamefont {Y.}~\bibnamefont {Liu}}, \bibinfo {author}
  {\bibfnamefont {G.-L.}\ \bibnamefont {Shen-Tu}}, \bibinfo {author}
  {\bibfnamefont {J.~S.}\ \bibnamefont {Pelc}}, \bibinfo {author}
  {\bibfnamefont {M.~M.}\ \bibnamefont {Fejer}}, \bibinfo {author}
  {\bibfnamefont {C.-Z.}\ \bibnamefont {Peng}}, \bibinfo {author}
  {\bibfnamefont {X.}~\bibnamefont {Ma}}, \bibinfo {author} {\bibfnamefont
  {Q.}~\bibnamefont {Zhang}}, \ and\ \bibinfo {author} {\bibfnamefont {J.-W.}\
  \bibnamefont {Pan}},\ }\href {\doibase 10.1103/PhysRevLett.114.180502}
  {\bibfield  {journal} {\bibinfo  {journal} {Phys. Rev. Lett.}\ }\textbf
  {\bibinfo {volume} {114}},\ \bibinfo {pages} {180502} (\bibinfo {year}
  {2015})}\BibitemShut {NoStop}%
\bibitem [{\citenamefont {Takesue}\ \emph {et~al.}(2015)\citenamefont
  {Takesue}, \citenamefont {Sasaki}, \citenamefont {Tamaki},\ and\
  \citenamefont {Koashi}}]{RRDPSexp1}%
  \BibitemOpen
  \bibfield  {author} {\bibinfo {author} {\bibfnamefont {H.}~\bibnamefont
  {Takesue}}, \bibinfo {author} {\bibfnamefont {T.}~\bibnamefont {Sasaki}},
  \bibinfo {author} {\bibfnamefont {K.}~\bibnamefont {Tamaki}}, \ and\ \bibinfo
  {author} {\bibfnamefont {M.}~\bibnamefont {Koashi}},\ }\href
  {http://dx.doi.org/10.1038/nphoton.2015.173} {\bibfield  {journal} {\bibinfo
  {journal} {Nat Photon}\ }\textbf {\bibinfo {volume} {9}},\ \bibinfo {pages}
  {827} (\bibinfo {year} {2015})}\BibitemShut {NoStop}%
\bibitem [{\citenamefont {{Wang}}\ \emph {et~al.}(2015)\citenamefont {{Wang}},
  \citenamefont {{Yin}}, \citenamefont {{Chen}}, \citenamefont {{He}},
  \citenamefont {{Song}}, \citenamefont {{Li}}, \citenamefont {{Zhang}},
  \citenamefont {{Zhou}}, \citenamefont {{Guo}},\ and\ \citenamefont
  {{Han}}}]{RRDPSexp2}%
  \BibitemOpen
  \bibfield  {author} {\bibinfo {author} {\bibfnamefont {S.}~\bibnamefont
  {{Wang}}}, \bibinfo {author} {\bibfnamefont {Z.-Q.}\ \bibnamefont {{Yin}}},
  \bibinfo {author} {\bibfnamefont {W.}~\bibnamefont {{Chen}}}, \bibinfo
  {author} {\bibfnamefont {D.-Y.}\ \bibnamefont {{He}}}, \bibinfo {author}
  {\bibfnamefont {X.-T.}\ \bibnamefont {{Song}}}, \bibinfo {author}
  {\bibfnamefont {H.-W.}\ \bibnamefont {{Li}}}, \bibinfo {author}
  {\bibfnamefont {L.-J.}\ \bibnamefont {{Zhang}}}, \bibinfo {author}
  {\bibfnamefont {Z.}~\bibnamefont {{Zhou}}}, \bibinfo {author} {\bibfnamefont
  {G.-C.}\ \bibnamefont {{Guo}}}, \ and\ \bibinfo {author} {\bibfnamefont
  {Z.-F.}\ \bibnamefont {{Han}}},\ }\href
  {http://dx.doi.org/10.1038/nphoton.2015.209} {\bibfield  {journal} {\bibinfo
  {journal} {Nat Photon}\ }\textbf {\bibinfo {volume} {9}},\ \bibinfo {pages}
  {832} (\bibinfo {year} {2015})}\BibitemShut {NoStop}%
\bibitem [{\citenamefont {Li}\ \emph {et~al.}(2016)\citenamefont {Li},
  \citenamefont {Cao}, \citenamefont {Dai}, \citenamefont {Lin}, \citenamefont
  {Zhang}, \citenamefont {Chen}, \citenamefont {Xu}, \citenamefont {Guan},
  \citenamefont {Liao}, \citenamefont {Yin}, \citenamefont {Zhang},
  \citenamefont {Ma}, \citenamefont {Peng},\ and\ \citenamefont
  {Pan}}]{RRDPSexp3}%
  \BibitemOpen
  \bibfield  {author} {\bibinfo {author} {\bibfnamefont {Y.-H.}\ \bibnamefont
  {Li}}, \bibinfo {author} {\bibfnamefont {Y.}~\bibnamefont {Cao}}, \bibinfo
  {author} {\bibfnamefont {H.}~\bibnamefont {Dai}}, \bibinfo {author}
  {\bibfnamefont {J.}~\bibnamefont {Lin}}, \bibinfo {author} {\bibfnamefont
  {Z.}~\bibnamefont {Zhang}}, \bibinfo {author} {\bibfnamefont
  {W.}~\bibnamefont {Chen}}, \bibinfo {author} {\bibfnamefont {Y.}~\bibnamefont
  {Xu}}, \bibinfo {author} {\bibfnamefont {J.-Y.}\ \bibnamefont {Guan}},
  \bibinfo {author} {\bibfnamefont {S.-K.}\ \bibnamefont {Liao}}, \bibinfo
  {author} {\bibfnamefont {J.}~\bibnamefont {Yin}}, \bibinfo {author}
  {\bibfnamefont {Q.}~\bibnamefont {Zhang}}, \bibinfo {author} {\bibfnamefont
  {X.}~\bibnamefont {Ma}}, \bibinfo {author} {\bibfnamefont {C.-Z.}\
  \bibnamefont {Peng}}, \ and\ \bibinfo {author} {\bibfnamefont {J.-W.}\
  \bibnamefont {Pan}},\ }\href {\doibase 10.1103/PhysRevA.93.030302} {\bibfield
   {journal} {\bibinfo  {journal} {Phys. Rev. A}\ }\textbf {\bibinfo {volume}
  {93}},\ \bibinfo {pages} {030302} (\bibinfo {year} {2016})}\BibitemShut
  {NoStop}%
\bibitem [{\citenamefont {Fuchs}\ \emph {et~al.}(2004)\citenamefont {Fuchs},
  \citenamefont {Schack},\ and\ \citenamefont {Scudo}}]{definetti}%
  \BibitemOpen
  \bibfield  {author} {\bibinfo {author} {\bibfnamefont {C.~A.}\ \bibnamefont
  {Fuchs}}, \bibinfo {author} {\bibfnamefont {R.}~\bibnamefont {Schack}}, \
  and\ \bibinfo {author} {\bibfnamefont {P.~F.}\ \bibnamefont {Scudo}},\ }\href
  {\doibase 10.1103/PhysRevA.69.062305} {\bibfield  {journal} {\bibinfo
  {journal} {Phys. Rev. A}\ }\textbf {\bibinfo {volume} {69}},\ \bibinfo
  {pages} {062305} (\bibinfo {year} {2004})}\BibitemShut {NoStop}%
\bibitem [{\citenamefont {Lo}\ \emph {et~al.}(2005{\natexlab{b}})\citenamefont
  {Lo}, \citenamefont {Chau},\ and\ \citenamefont
  {Ardehali}}]{Lo:EffBB84:2005}%
  \BibitemOpen
  \bibfield  {author} {\bibinfo {author} {\bibfnamefont {H.-K.}\ \bibnamefont
  {Lo}}, \bibinfo {author} {\bibfnamefont {H.~F.}\ \bibnamefont {Chau}}, \ and\
  \bibinfo {author} {\bibfnamefont {M.}~\bibnamefont {Ardehali}},\ }\href@noop
  {} {\bibfield  {journal} {\bibinfo  {journal} {Journal of Cryptology}\
  }\textbf {\bibinfo {volume} {18}},\ \bibinfo {pages} {133} (\bibinfo {year}
  {2005}{\natexlab{b}})}\BibitemShut {NoStop}%
\bibitem [{\citenamefont {He}\ \emph {et~al.}(2017)\citenamefont {He},
  \citenamefont {Wang}, \citenamefont {Chen}, \citenamefont {Yin},
  \citenamefont {Qian}, \citenamefont {Zhou}, \citenamefont {Guo},\ and\
  \citenamefont {Han}}]{he2017sine}%
  \BibitemOpen
  \bibfield  {author} {\bibinfo {author} {\bibfnamefont {D.-Y.}\ \bibnamefont
  {He}}, \bibinfo {author} {\bibfnamefont {S.}~\bibnamefont {Wang}}, \bibinfo
  {author} {\bibfnamefont {W.}~\bibnamefont {Chen}}, \bibinfo {author}
  {\bibfnamefont {Z.-Q.}\ \bibnamefont {Yin}}, \bibinfo {author} {\bibfnamefont
  {Y.-J.}\ \bibnamefont {Qian}}, \bibinfo {author} {\bibfnamefont
  {Z.}~\bibnamefont {Zhou}}, \bibinfo {author} {\bibfnamefont {G.-C.}\
  \bibnamefont {Guo}}, \ and\ \bibinfo {author} {\bibfnamefont {Z.-F.}\
  \bibnamefont {Han}},\ }\href@noop {} {\bibfield  {journal} {\bibinfo
  {journal} {Applied Physics Letters}\ }\textbf {\bibinfo {volume} {110}},\
  \bibinfo {pages} {111104} (\bibinfo {year} {2017})}\BibitemShut {NoStop}%
\bibitem [{\citenamefont {Sasaki}\ \emph {et~al.}(2016)\citenamefont {Sasaki},
  \citenamefont {Tamaki},\ and\ \citenamefont {Koashi}}]{sasaki2016quantum}%
  \BibitemOpen
  \bibfield  {author} {\bibinfo {author} {\bibfnamefont {T.}~\bibnamefont
  {Sasaki}}, \bibinfo {author} {\bibfnamefont {K.}~\bibnamefont {Tamaki}}, \
  and\ \bibinfo {author} {\bibfnamefont {M.}~\bibnamefont {Koashi}},\
  }\href@noop {} {\bibfield  {journal} {\bibinfo  {journal} {arXiv preprint
  arXiv:1604.04460}\ } (\bibinfo {year} {2016})}\BibitemShut {NoStop}%
\end{thebibliography}%

\end{document}